\documentclass[amsmath,amssymb,prc,final,showpacs,twocolumn]{revtex4-1}
 
\usepackage{bm,hyphenat,xspace}
\usepackage{graphicx,epsfig}

\newcommand{\Refs}{Refs.}

\newcommand {\mct}{\mathcal{T}}

\newcommand{\He}{{}^3\mathrm{He}}
\newcommand{\Hh}{{}^3\mathrm{H}}

\newcommand{\pH}{p\text{-}{}^3\mathrm{H}}
\newcommand{\nHe}{n\text{-}{}^3\mathrm{He}}

\newcommand{\dd}{d\text{-}d}

\begin{document}

\title {Polarization observables and spin-aligned fusion rates in
${}^2\mathrm{H}(d,p){}^3\mathrm{H}$ and ${}^2\mathrm{H}(d,n){}^3\mathrm{He}$ reactions}
  
\author{A.~Deltuva} 
\email{deltuva@cii.fc.ul.pt}
\author{A.~C.~Fonseca} 
\affiliation{Centro de F\'{\i}sica Nuclear da Universidade de Lisboa, 
P-1649-003 Lisboa, Portugal }

\received{ February 24, 2010}

\begin{abstract}
Nucleon transfer reactions in low-energy deuteron-deuteron scattering
are described by solving
exact four-particle equations in momentum space.
The Coulomb interaction between the protons is included using the screening 
and renormalization method.
Various realistic potentials are used between nucleon pairs.
The energy dependence of the differential cross section,
analyzing powers, polarizations, spin-transfer coefficient,
and the quintet suppression factor is studied.
\end{abstract}

 \maketitle


Neutron ($n$) and proton ($p$) transfer reactions in
deuteron-deuteron ($\dd$) scattering, $d+d \to p+\Hh$ and  $d+d \to n+\He$,
 are amongst the simplest nuclear reactions where charge-symmetry breaking (CSB) in the nuclear force 
can be verified or searched for. For that purpose one needs a very precise calculation of the 
four-nucleon ($4N$) problem with different interaction models based on nucleon-nucleon ($NN$) 
and many-nucleon forces
 together with the inclusion of the Coulomb force between protons which is the most important cause 
of CSB. Such task is now possible in view of the progress achieved in the past few 
years \cite{deltuva:07a,deltuva:07c} on the solution 
of exact Alt, Grassberger, and Sandhas (AGS) equations \cite{grassberger:67}
for four-particle transitions operators that,
in addition to the strong nuclear force,  include also the Coulomb  interaction.

The aim of the present paper is to study the energy dependence of the  $d+d \to p+\Hh$ and  $d+d \to n+\He$
 observables below three-body breakup threshold using different nuclear force models. 
In addition to the differential cross sections \cite{blair:48b,hunter:49,gruebler:72a,gruebler:81a}
there are precise measurements of the deuteron 
analyzing powers between 1.5 and 4 MeV deuteron lab energy \cite{gruebler:72a,dries:79a}.
The  analyzing powers show a complex structure of maxima and minima that in some cases varies 
rapidly with the energy. These rapid variations and complex structure means that different $dd$ 
partial waves play a significant role, and agreement or disagreement with experimental data at 
different energies may result from a delicate interplay between them. The contribution of each 
partial wave to the observables and their evolution with the energy is analyzed.

Other observables that were measured but
 never analyzed through exact $4N$ calculations are 
polarization of the outgoing nucleon \cite{ddNBPy}
and deuteron to nucleon spin transfer coefficients
\cite{lisowski:75,ddnhkyy:tunl}; they are also calculated in the present work.

Furthermore, we will get back to an old issue that remains of interest to $\dd$ fusion in hot plasma. 
It was initially believed \cite{kulsrud:82}
that at very low energy the production of neutrons resulting from 
$d+d \to n+\He$ could be controlled by polarizing all deuterons in the plasma such that only 
spin-aligned $\dd$ fusion would take place. If spin-aligned $d+d \to n+\He$ reaction would be 
significantly suppressed compared to $d+d \to p+\Hh$,
one would have a significant reduction of undesired neutrons coming out of the plasma. 
In the past many approximate calculations were made that 
yielded contradicting results. More sophisticated calculations  \cite{hofmann:84a}
indicated that no strong suppression should be expected. The present well-converged
$4N$ calculations aim at settling
the assumption on the possible suppression of $d+d \to n+\He$ reaction with spin-aligned  deuterons.

Our description of the $4N$ system is based on
exact four-particle equations for the transition operators
as derived by Alt, Grassberger, and Sandhas \cite{grassberger:67};
they are equivalent to the Faddeev-Yakubovsky equations \cite{yakubovsky:67}
for the wave-function components.
Since in the isospin formalism  protons and neutrons can be considered
as identical particles, the symmetrized form of the AGS equations 
\cite{deltuva:07a,deltuva:07c} is appropriate.
Although the initial $\dd$ state has total isospin $\mct = 0$,
the final $\nHe$ and $\pH$ states are dominated by 
both isospin $\mct = 0$ and $\mct = 1$ components;
a very small admixture of $\mct = 2$ is present due to the
charge dependence of the hadronic and electromagnetic interactions. 
The most important source for the total isospin nonconservation
is the Coulomb force between the protons; it is included
using the method of screening and renormalization
\cite{taylor:74a,alt:80a,deltuva:05a,deltuva:07b}.
The hadronic charge dependence, as given by the modern $NN$ potentials,
is taken into account as well.
After partial wave decomposition, the
AGS equations are a system of three-variable integral equations that are solved
numerically without any approximation beyond the usual discretization
of momentum variables on a finite mesh;
technical details are given in Refs.~\cite{deltuva:07a,deltuva:07c}.
The results we present are well converged with respect to the partial-wave
expansion, the number of momentum meshpoints, and the Coulomb screening radius
used to calculate the short-range part of the amplitudes.
In this way the discrepancies with the experimental data may be attributed solely to 
the underlying $NN$ forces or lack of many-nucleon forces.

As two-nucleon interactions we use the phenomenological potentials
Argonne V18 (AV18, hadronic part only)~\cite{wiringa:95a},
charge-dependent Bonn (CD Bonn)~\cite{machleidt:01a}, and
inside nonlocal outside Yukawa (INOY04) potential by
Doleschall~\cite{doleschall:04a}, and the one derived from the chiral
perturbation theory at next-to-next-to-next-to-leading order
(N3LO)~\cite{entem:03a}.
Furthermore, we consider also a two-baryon coupled-channel potential
including virtual excitation of a nucleon to a $\Delta$ isobar \cite{deltuva:03c}
that in the $4N$ system yields effective $3N$ and $4N$ forces \cite{deltuva:08a}.
Point Coulomb is added for $pp$ pairs.
 $\He$ and $\Hh$  binding energies calculated with those potentials are
collected in  the Table \ref{tab:1}. Thus,
the  presence of the $3N$ force is also simulated
by the potential INOY04 that fits both $\He$ and $\Hh$ experimental 
binding energies  and thereby 
of all the used potentials is the only one that reproduces correctly
the momenta of the final $\nHe$ and $\pH$ states. 
For this reason it is not surprising to see that INOY04
 potential gives the best description of $d+d \to
n+\He$ and $d+d \to p+\Hh$ data. We therefore show the predictions of all 
potentials for the reactions at deuteron lab energy $E_d = 3$ MeV 
but study the energy dependence of the observables using INOY04 only.

\begin{table}[htbp]
\begin{ruledtabular}
\begin{tabular}{l*{3}{c}}
 & $B(\Hh)$ & $B(\He)$ & $P_D(d)$   \\  \hline
AV18        & 7.66 & 6.95 & 5.78 \\
N3LO        & 7.85 & 7.13 & 4.51 \\
CD Bonn     & 8.00 & 7.26 & 4.85 \\
CD Bonn + $\Delta$  & 8.28 & 7.53 & 4.85 \\
INOY04      & 8.49 & 7.73 & 3.60 \\
Experiment  & 8.48 & 7.72 & 
\end{tabular}
\end{ruledtabular}
\caption{$\Hh$ and $\He$ binding energies (in MeV)
and deuteron  $D$-state probability  $P_D(d)$ (in percent)
for different $NN$ potentials.}
\label{tab:1}
\end{table}

\begin{figure}[!]
\begin{center}
\includegraphics[scale=0.55]{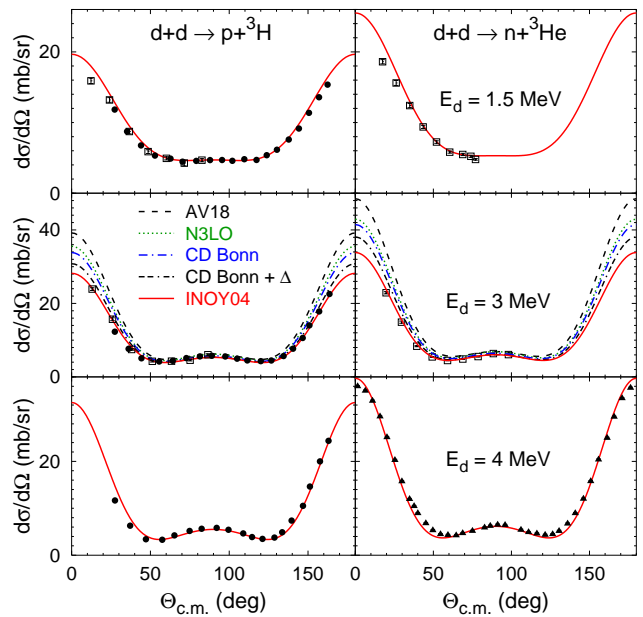}
\end{center}
\caption{\label{fig:S} (Color online)
Differential cross section of 
$d+d \to p+\Hh$ and  $d+d \to n+\He$ reactions at 1.5, 3, and 4 MeV deuteron lab energy
as function of the nucleon c.m. scattering angle. Results obtained with various realistic
$NN$ potentials are compared with 
the experimental data   from \Refs~\cite{blair:48b} (squares), 
 \cite{gruebler:72a,gruebler:81a} (circles), and \cite{hunter:49}  (triangles);
the latter set is taken at $E_d = 3.7$ MeV.}
\end{figure}

Although $\dd$ elastic scattering is calculated simultaneously with the transfer reactions,
it is a less interesting case and therefore 
not discussed in the present work: As demonstrated in Refs.~\cite{deltuva:07c,deltuva:08a},
the $\dd$ elastic cross section data is well described by all $NN$ force models and the corresponding
deuteron analyzing powers are very small with quite large error bars that preclude physics conclusions.

In Fig.~\ref{fig:S} we present the differential cross section 
$d\sigma/d\Omega$ results
for both $d+d \to p+\Hh$ and  $d+d \to n+\He$ reactions at $E_D = 1.5$, 3, 
and 4 MeV as function of the nucleon center-of-mass (c.m.) scattering angle
$\Theta_{\mathrm{c.m.}}$.
 This observable is symmetric with respect to $\Theta_{\mathrm{c.m.}} = 90^\circ$.
While at the lowest energy the differential cross section is flat
between $\Theta_{\mathrm{c.m.}} = 60^\circ$ and $120^\circ$, with increasing energy
it develops a minima around  $\Theta_{\mathrm{c.m.}} = 55^\circ$ and $125^\circ$
and a local maximum at $\Theta_{\mathrm{c.m.}} = 90^\circ$.
Furthermore,  $d\sigma/d\Omega$ increases with energy
at forward and backward angles. There is a good agreement between
the experimental data and the predictions of the
INOY04 potential whereas other models overestimate the data, especially
 at  forward and backward angles; furthermore, the 
discrepancy seems to be proportional to the defect in theoretical $\He$ and $\Hh$
binding energies. Thus, the differential cross section in the considered
transfer reactions correlates nearly linearly with $3N$ binding energies.

\begin{figure}[!]
\begin{center}
\includegraphics[scale=0.55]{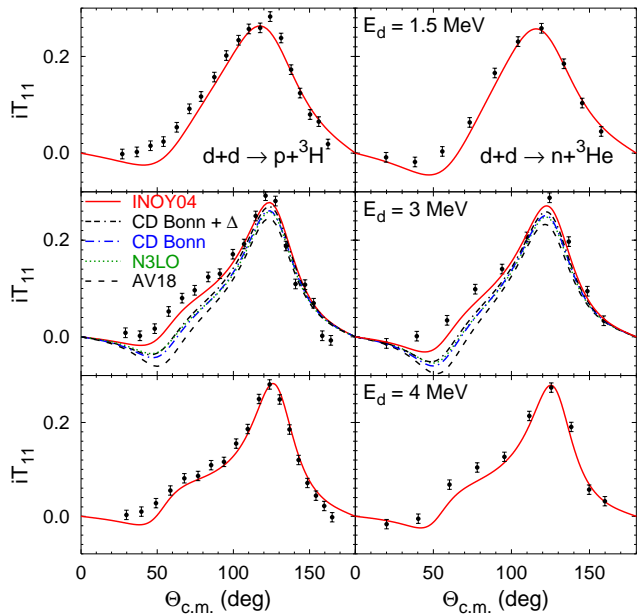}
\end{center}
\caption{\label{fig:T11} (Color online)
Deuteron vector analyzing power $iT_{11}$ of 
$d+d \to p+\Hh$ and  $d+d \to n+\He$ reactions at 1.5, 3, and 4 MeV deuteron lab energy. 
The data are from Ref.~\cite{gruebler:72a,gruebler:81a} for $d+d \to p+\Hh$ and from
Ref.~\cite{dries:79a} for $d+d \to n+\He$.}
\end{figure}

\begin{figure}[!]
\begin{center}
\includegraphics[scale=0.55]{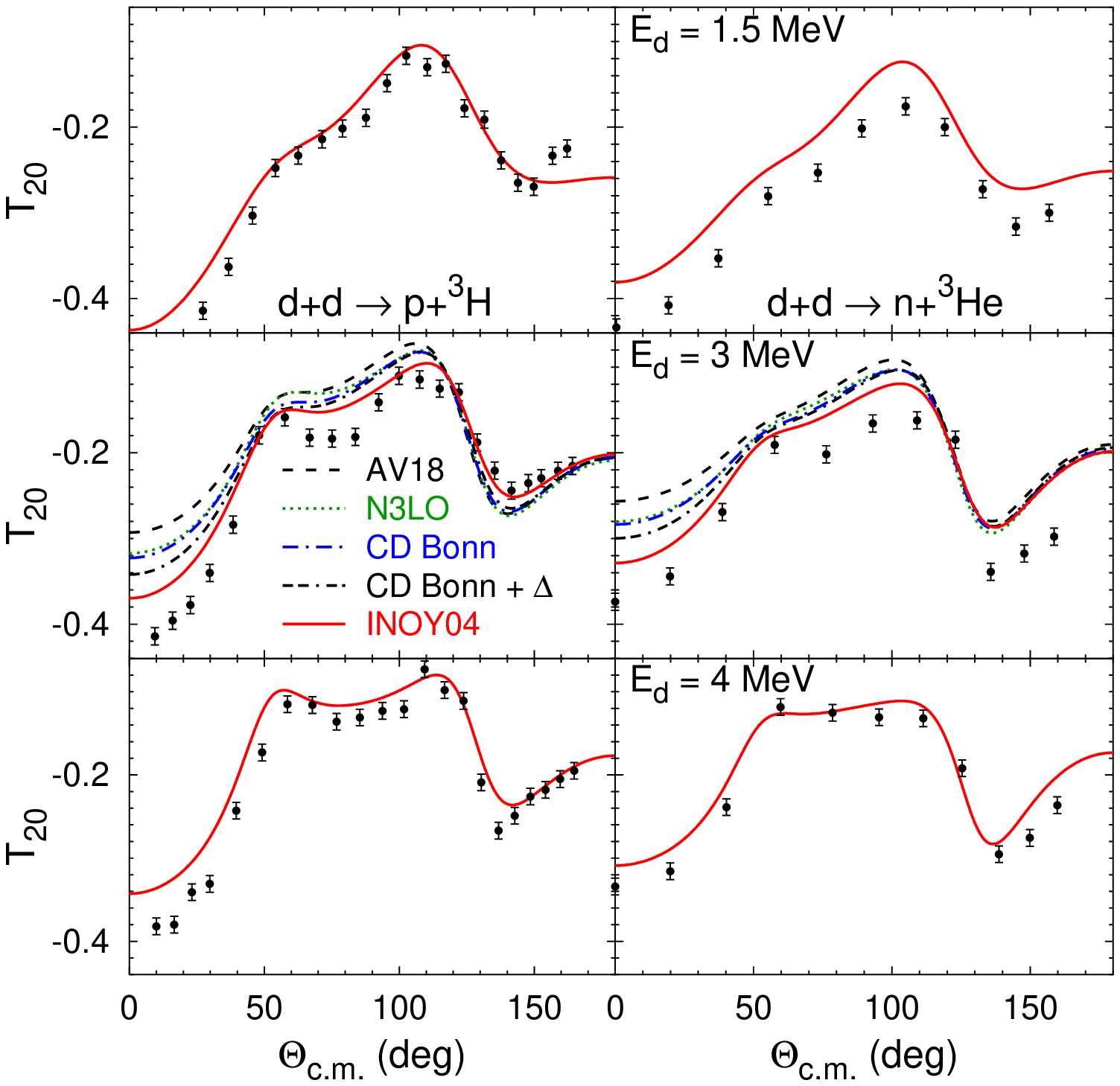}
\end{center}
\caption{\label{fig:T20} (Color online)
Same as Fig.~\ref{fig:T11} but for the
deuteron tensor analyzing power $T_{20}$.}
\end{figure}

\begin{figure}[!]
\begin{center}
\includegraphics[scale=0.55]{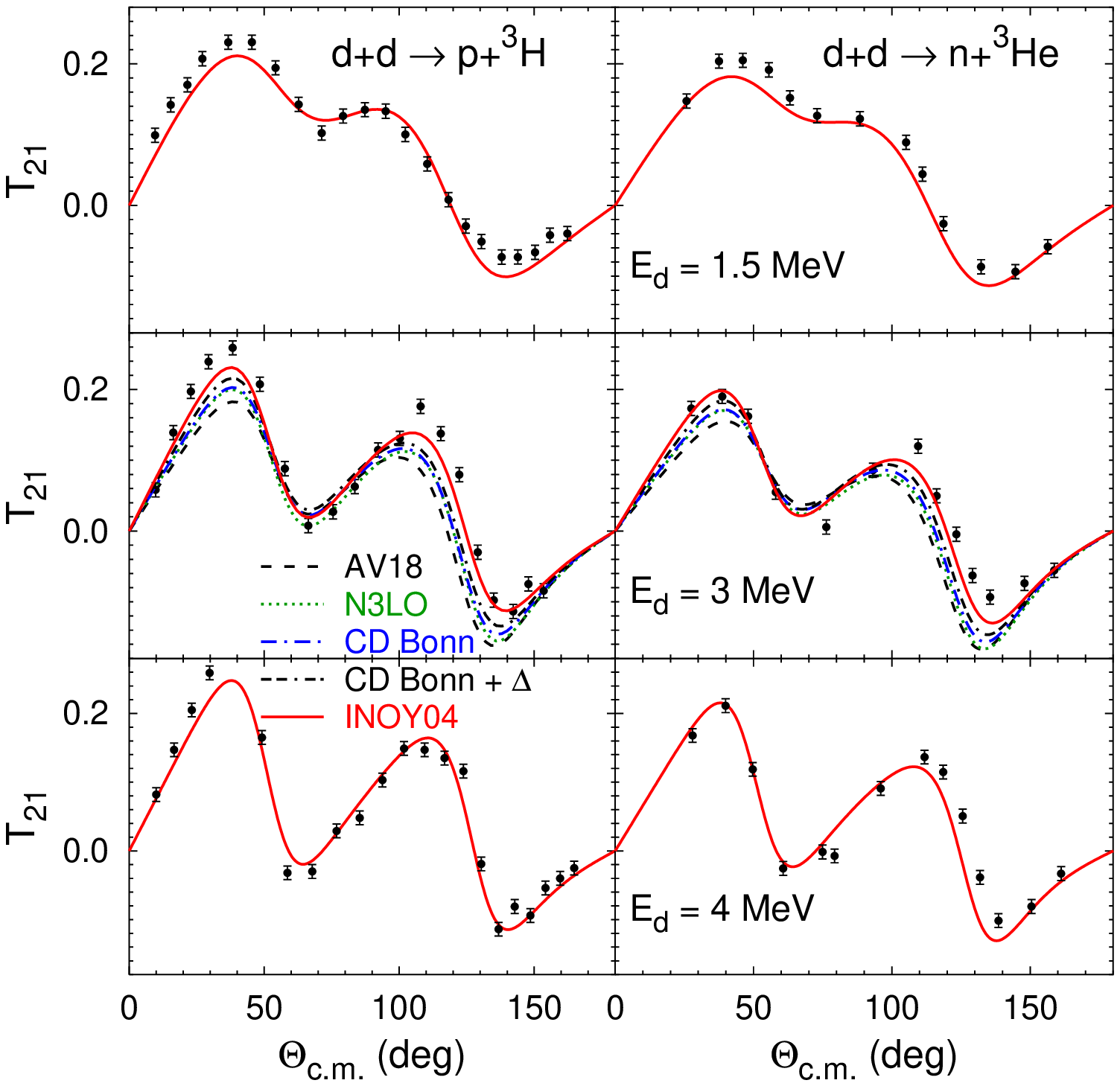}
\end{center}
\caption{\label{fig:T21} (Color online)
Same as Fig.~\ref{fig:T11} but for the
deuteron tensor analyzing power $T_{21}$.}
\end{figure}

\begin{figure}[!]
\begin{center}
\includegraphics[scale=0.55]{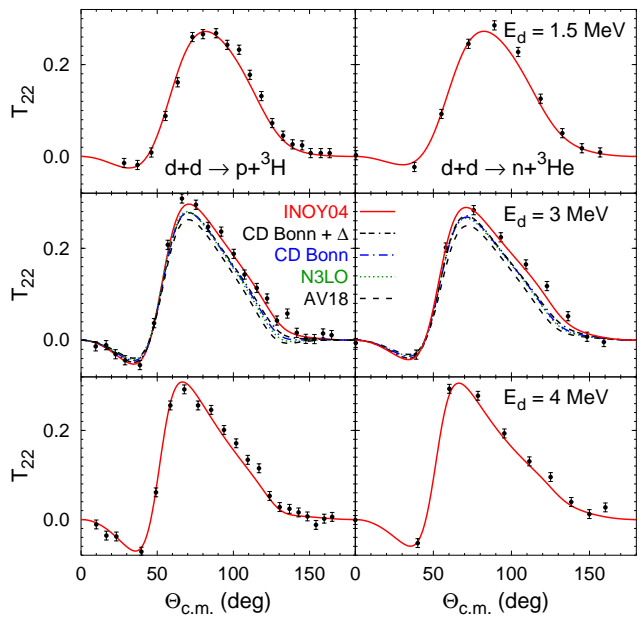}
\end{center}
\caption{\label{fig:T22} (Color online)
Same as Fig.~\ref{fig:T11} but for the
deuteron tensor analyzing power $T_{22}$.}
\end{figure}

In Figs.~\ref{fig:T11} - \ref{fig:T22} we present the corresponding results
for the deuteron analyzing powers.
The vector analyzing power $iT_{11}$ in Fig.~\ref{fig:T11} 
varies quite slowly with the energy. The experimental data
are well reproduced by the INOY04 model, although a small discrepancy
remains around $\Theta_{\mathrm{c.m.}} = 45^\circ$.
More significant discrepancies take place in the tensor analyzing power $T_{20}$
(Fig.~\ref{fig:T20})
 at small scattering angles, although this observable is slightly overestimated
almost in the whole kinematical regime. Nevertheless, the energy dependence
of  $T_{20}$, being considerably stronger than for $iT_{11}$, is well reproduced
by the calculations. The tensor analyzing power $T_{21}$ (Fig.~\ref{fig:T21}),
though showing strong energy dependence at intermediate angles, is quite well accounted for
by the INOY04 predictions. The  description of the tensor analyzing power $T_{22}$ 
(Fig.~\ref{fig:T22}) is even better; this observable, much like $iT_{11}$,
varies slowly with the energy.
The results obtained with other $NN$ potentials at $E_d = 3$ MeV deviate
from the respective data more significantly. Furthermore,  even if the reproduction
of experimental  binding energies is an important factor, it is certainly not the only 
one that matters since the linear correlation with 
$3N$ binding energies seems to be violated, e.g., for $iT_{11}$ at $\Theta_{\mathrm{c.m.}}$
between $40^\circ$  and $90^\circ$, or for $T_{22}$ between $\Theta_{\mathrm{c.m.}} = 60^\circ$ 
and $120^\circ$,
the predictions of N3LO, CD Bonn, and CD Bonn + $\Delta$ almost coincide
despite quite significant differences (up to 0.4 MeV) in the respective $3N$ 
binding energies while the predictions of INOY04 and AV18 stay well separated.
A closer look into the calculated properties of deuteron, $\He$, and $\Hh$
as given in Table \ref{tab:1} suggests that the considered spin observables
correlate, in addition to the $3N$ binding  energy, also with
the $D$-state probability of deuteron $P_D(d)$.
The correlation between $P_D(d)$ and  $3N$ binding  energy takes place
for most phenomenological $NN$ potentials, but  N3LO and CD Bonn + $\Delta$
models clearly violate that correlation thereby allowing to study
the dependence of $4N$ observables on both $3N$ binding  energy and  $P_D(d)$.
The results in Figs.~\ref{fig:T11} - \ref{fig:T22} indicate that increasing the $3N$ binding energy 
and decreasing $P_D(d)$ move the theoretical predictions into the same direction 
but the corresponding rates, i.e., strength of the correlations, depend
on the observable and kinematical regime. One may conjecture that
by a slight decrease of $P_D(d)$ of INOY04 while keeping the $\He$ and $\Hh$
 binding  energies unchanged one might be able to cure the
$iT_{11}$ discrepancy around $\Theta_{\mathrm{c.m.}} = 45^\circ$. However,
$T_{20}$ shows quite weak correlation with  $P_D(d)$ and therefore its
description would not be improved significantly by a small decrease of $P_D(d)$.

\begin{figure}[!]
\begin{center}
\includegraphics[scale=0.55]{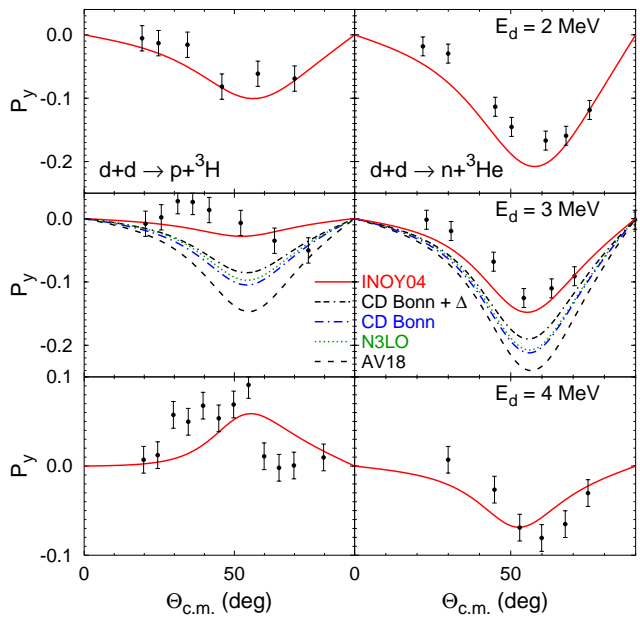}
\end{center}
\caption{\label{fig:P} (Color online)
Outgoing nucleon polarization $P_y$ of
$d+d \to p+\Hh$ and  $d+d \to n+\He$ reactions at 2, 3, and 4 MeV deuteron lab energy. 
The data are from Ref.~\cite{ddNBPy}.}
\end{figure}

\begin{figure}[!]
\begin{center}
\includegraphics[scale=0.65]{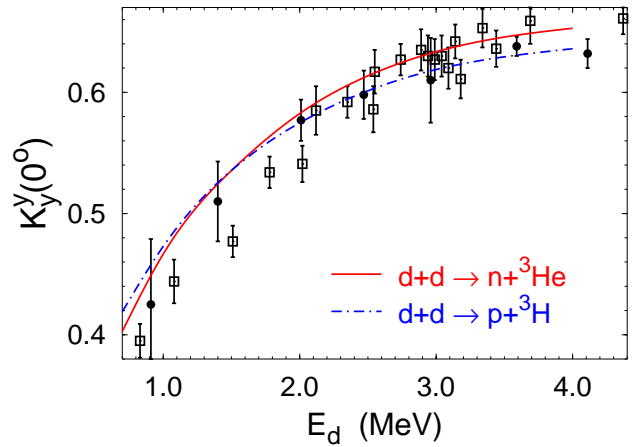}
\end{center}
\caption{\label{fig:kyy} (Color online)
Deuteron to nucleon spin transfer coefficient $K_y^y$ 
for  $d+d \to p+\Hh$ and  $d+d \to n+\He$ reactions at
 $\Theta_{\mathrm{c.m.}} = 0^\circ$ as function of the deuteron lab energy.
The data are from Ref.~\cite{lisowski:75} (circles) and  from
Ref.~\cite{ddnhkyy:tunl} (squares); both sets refer to the  $d+d \to n+\He$ reaction.}
\end{figure}

In Fig.~\ref{fig:P} we present the results for the polarization $P_y$ of the outgoing nucleon in 
$d+d \to p+\Hh$ and  $d+d \to n+\He$ reactions; this observable is equivalent
to the nucleon analyzing power $A_y$ in the time-reverse reactions
$p+\Hh \to d+d$ and  $n+\He \to d+d$. $P_y$ is antisymmetric with respect to 
$\Theta_{\mathrm{c.m.}} = 90^\circ$; we therefore show
only the angular regime up to $\Theta_{\mathrm{c.m.}} = 90^\circ$ that contains all the available data.
$P_y$ shows quite strong energy dependence; in the $d+d \to p+\Hh$ case it even changes 
the sign when $E_d$ varies from 2 MeV to 4 MeV being almost zero at $E_d = 3$ MeV.
In the $d+d \to n+\He$ case the observable changes with the same trend but with about
1.5 MeV shift in the energy.
The qualitative reproduction of the experimental data having large errorbars is quite
successful by the potential INOY04 whereas predictions of the other interaction models 
are further away. As the deuteron analyzing powers, the nucleon polarization correlates
not only with the $3N$ binding  energy but also with $P_D(d)$ as one can see by comparing
N3LO, CD Bonn, and CD Bonn + $\Delta$ results.

Next we consider double polarization observables for which, unfortunately,
 the experimental data are much scarcer. 
The deuteron to neutron spin transfer coefficient $K_y^y$ 
has been measured in several experiments \cite{lisowski:75,ddnhkyy:tunl}, however, only for the 
neutrons emitted in the forward direction  $\Theta_{\mathrm{c.m.}} = 0^\circ$;
the data for the corresponding observable in the $d+d \to p+\Hh$ reaction
is not available below the three-body breakup threshold.
In Fig.~\ref{fig:kyy} we compare the 
deuteron to neutron spin transfer coefficient $K_y^y(0^\circ)$ 
calculated using the INOY04 potential with two sets of experimental data
in the energy range  $0.75 \le E_d \le 4$ MeV.
There is a good agreement above $E_d = 2$ MeV while at lower energies,
where the observable shows stronger energy dependence, the data points
from Ref.~\cite{ddnhkyy:tunl}  are slightly overpredicted.
We include in  Fig.~\ref{fig:kyy} also the calculated $K_y^y(0^\circ)$  for the
$d+d \to p+\Hh$ reaction that shows a similar behavior. 
The dependence on the $NN$ force model is quite weak for  $K_y^y(0^\circ)$,
considerably smaller than the experimental error bars:
at  $E_d = 3$ MeV the predictions of AV18, N3LO, CD Bonn, CD Bonn + $\Delta$, and INOY04
are 0.622, 0.626, 0.630, 0.632, and 0.633, respectively. 

\begin{figure}[!]
\begin{center}
\includegraphics[scale=0.55]{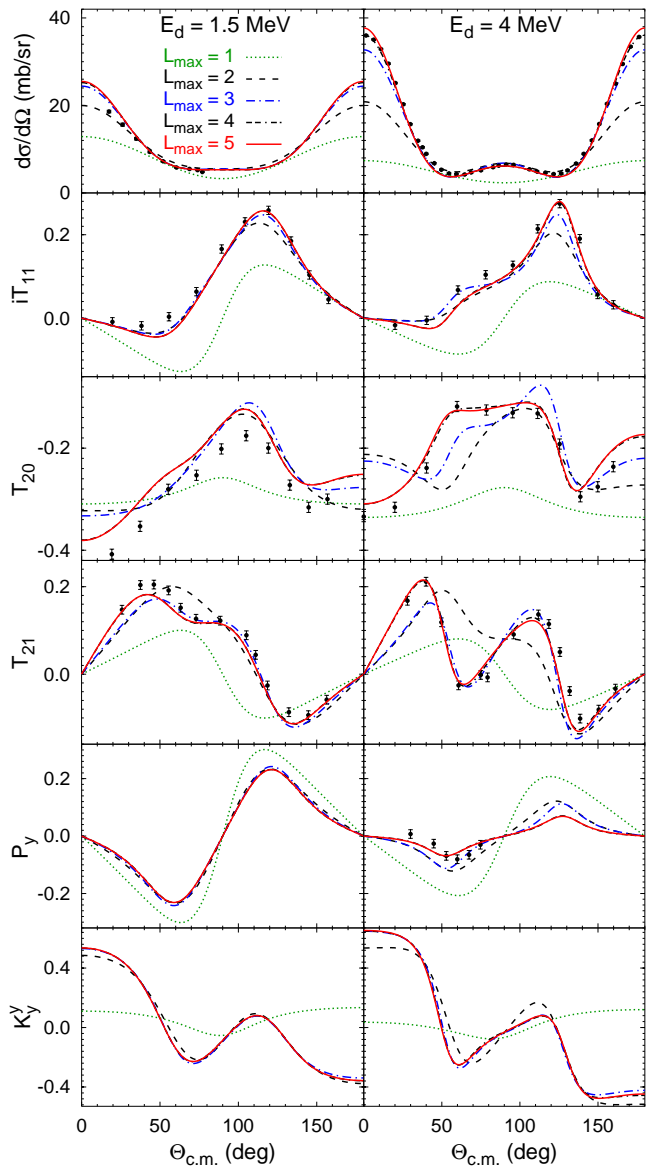}
\end{center}
\caption{\label{fig:pw} (Color online)
Observables of $d+d \to n+\He$ reaction at $E_d = 1.5$ and 4 MeV.
Results including initial and final states with 
two-cluster relative orbital angular momentum  $ L \le L_{\mathrm{max}}$ 
are compared for $L_{\mathrm{max}}$ ranging from 1 to 5.
$L_{\mathrm{max}} = 4$ and 5 results lie almost on top of each other.
The experimental data  are as in previous figures.}
\end{figure}

\begin{figure}[!]
\begin{center}
\includegraphics[scale=0.65]{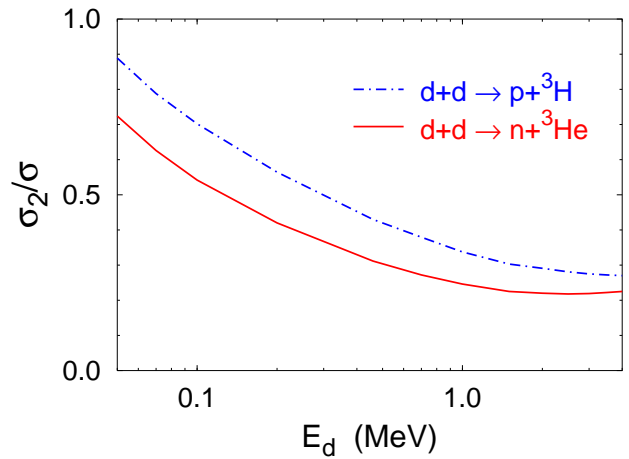}
\end{center}
\caption{\label{fig:qsf} (Color online)
The quintet suppression factor $\sigma_2/\sigma$
for  $d+d \to p+\Hh$ and  $d+d \to n+\He$ reactions 
as function of the deuteron lab energy.}
\end{figure}

In Fig.~\ref{fig:pw} we study the contribution of various initial and final state
partial waves characterized by the relative two-cluster, i.e.,
$\dd$, $\nHe$, and $\pH$, orbital angular momentum $L$.
Since $d+d \to p+\Hh$ and  $d+d \to n+\He$ observables show similar behavior,
we restrict our study to the latter reaction at $E_d = 1.5$ and 4 MeV. 
We performed a series of calculations with INOY04 potential
including $L \le L_{\mathrm{max}}$ with $L_{\mathrm{max}}$ ranging from 1 to 5.
The predictions with  $L_{\mathrm{max}} = 4$ and 5 are very close, indicating that
the $L=5$ contribution is very small;  however, since the convergence
is not monotonic, we have also verified that $L=6$ contribution can be safely neglected.
In contrast, the partial waves with $L=4$  yield a sizable contribution at $E_d = 4$ MeV
and for the deuteron tensor analyzing powers even at  $E_d = 1.5$ MeV. 
For other observables at this energy  $L_{\mathrm{max}} = 3$ is sufficient 
while  $P_y$  is reasonably well converged with  $L_{\mathrm{max}} = 2$.
It is interesting to note that $L_{\mathrm{max}} = 1$ results for most observables
vary slowly with the energy while stronger energy dependence 
comes from $L=2$, 3, and 4 partial waves. 
The above analysis partially explains sizable differences between our converged results
and those of Ref.~\cite{hofmann:07a} obtained using the resonanting group method (RGM) since
the latter work included only $L \le 3$ states.
Our AV18 results calculated with $L_{\mathrm{max}} = 3$ (not shown here) are qualitatively similar
to the corresponding results of  Ref.~\cite{hofmann:07a}.

Finally we present results for the observables characterizing the spin correlations
of the initial-state deuterons. We are not aware of the existence of the experimental
data, but there are plans \cite{schieck:pc,didelez:pc}
for future experiments involving the direct measurement of the
so-called quintet suppression factor (QSF) defined as the ratio $\sigma_2/\sigma$,
where $\sigma_2$ is the cross section for the considered transfer reaction with the 
spins of the initial state deuterons being parallel,  i.e., with the total spin being 2,  
and $\sigma$ is the unpolarized (spin-averaged) cross section.
As mentioned in the introduction, this observable is relevant for $\dd$ fusion in hot plasma.
Results obtained using INOY04 potential for $E_d$ between 50 keV and 4 MeV
are shown in Fig.~\ref{fig:qsf}.
The QSF for the $d+d \to n+\He$ reaction is indeed smaller than for 
$d+d \to p+\Hh$; however, the difference is only about 20 - 25\% in the whole considered regime. 
The energy dependence of the QSF is in both reactions similar: it is of the order of 0.25 above
$E_d = 2$ MeV but increases more rapidly  as the energy decreases.
Nevertheless, our QSF prediction are somehow smaller than the values obtained from
 $R$-matrix analysis and RGM \cite{hofmann:84a}.
This observable shows also strong dependence on the $NN$ force model, e.g.,
the QSF values for the $d+d \to n+\He$ reaction at $E_d = 3$ MeV  predicted by 
AV18, N3LO, CD Bonn, CD Bonn + $\Delta$, and INOY04 models
are 0.133, 0.159, 0.164, 0.185, and 0.219,  respectively. Thus, the QSF  correlates with 
both  $3N$ binding  energy and $P_D(d)$, increasing when the former increases and/or the latter 
decreases. 

In summary,
we performed exact four-particle calculations of 
$d+d \to p+\Hh$ and  $d+d \to n+\He$ reactions with several realistic $NN$ potentials.
Energy dependence of the differential cross section and spin observables was studied
below three-body breakup threshold. Correlations of the predictions
with $3N$ binding  energy and deuteron  $D$-state probability  $P_D(d)$ 
were observed. The INOY04 potential that fits
both $\He$ and $\Hh$ experimental binding energies and has 
the smallest  $P_D(d) = 3.60 \%$  among all realistic potentials,  accounts well for 
the experimental data with only few discrepancies, e.g., the one in the 
deuteron tensor analyzing power $T_{20}$. Some data suggest that even smaller
 $P_D(d)$ would be preferred.
We also predict the quintet suppression factor in the  $d+d \to n+\He$ reaction to be
only up to 25\% smaller than the one in $d+d \to p+\Hh$; the 
$\dd$ fusion with spin-aligned  deuterons seems to be significantly suppressed at few MeV
but not in the keV regime.




\end{document}